\documentclass[aps,prb,reprint,superscriptaddress]{revtex4-2}

\usepackage{bm}
\usepackage{graphicx}
\usepackage[usenames, dvipsnames]{color}
\usepackage{dcolumn}
\usepackage{float}
\usepackage{hyperref}

\newcommand{\rw}{\color{black}}

\begin{document}
\title{MIASANS at the longitudinal neutron resonant spin-echo spectrometer RESEDA}
%
%

\author{Jonathan C. Leiner}
\email[]{Corresponding author: jon.leiner@tum.de}
\affiliation{Physik-Department, Technische Universit\"{a}t M\"{u}nchen, D-85748 Garching, Germany}
\affiliation{Heinz Maier-Leibnitz Zentrum (MLZ), Technische Universit\"at M\"unchen, D-85748 Garching, Germany}

\author{Christian Franz}
\affiliation{J\"ulich Centre for Neutron Science JCNS, Forschungszentrum J\"ulich GmbH Outstation at MLZ FRM\,-\,II, D-85747 Garching, Germany}

\author{Johanna K. Jochum}
\affiliation{Physik-Department, Technische Universit\"{a}t M\"{u}nchen, D-85748 Garching, Germany}
\affiliation{Heinz Maier-Leibnitz Zentrum (MLZ), Technische Universit\"at M\"unchen, D-85748 Garching, Germany}

\author{Christian Pfleiderer}
\affiliation{Physik-Department, Technische Universit\"{a}t M\"{u}nchen, D-85748 Garching, Germany}
\affiliation{Center for Quantum Engineering (ZQE), Technische Universit\"at M\"unchen, D-85748 Garching, Germany}
\affiliation{Munich Center for Quantum Science and Technology (MCQST), Technische Universit\"at M\"unchen, D-85748 Garching, Germany}



\begin{abstract}
  The RESEDA (Resonant Spin-Echo for Diverse Applications) instrument has been optimized for neutron scattering measurements of quasi-elastic and inelastic processes over a wide parameter range.  One spectrometer arm of RESEDA is configured for the MIEZE (Modulation of Intensity with Zero Effort) technique, where the measured signal is an oscillation in neutron intensity over time prepared by two precisely tuned radio-frequency (RF) flippers. With MIEZE, all spin-manipulations are performed before the beam reaches the sample, and thus the signal from sample scattering is not disrupted by any depolarizing conditions there (i.e. magnetic materials and fields). The MIEZE spectrometer is being further optimized for the requirements of small angle neutron scattering (MIASANS), a versatile combination of the spatial and dynamical resolving power of both techniques. We present the current status of (i) the newly installed superconducting solenoids as part of the RF flippers to significantly extend the dynamic range (ii) the development and installation of a new detector on a translation stage within a new larger SANS-type vacuum vessel for flexibility with angular coverage and resolution, and (iii) the efforts to reduce background.
\end{abstract}
\maketitle
\section{Introduction}
\label{intro}

\subsection{Motivation}
\label{motivation}

Small angle neutron scattering (SANS) represents an exceptionally important probe for a wide range of scientific problems. SANS is a powerful tool for studying the structure of complex fluids and polymers \cite{C1SM06257C, D0SM01962C}. SANS is also routinely practical for studying the soft matter materials and structures such as biological macromolecules, e.g. proteins \cite{perkins1988structural} and DNA \cite{kennaway2012structure}. In addition, SANS is particularly useful in resolving mesoscale magnetic structures \cite{2019Kindervater} such as those of magnetic nanoparticles \cite{D1NA00482D, Bender2022} and vortices in superconductors \cite{Eskildsen_2011}. The comprehensive utilities of SANS are difficult to fully enumerate, as shown by the breadth and depth contained in various review articles \cite{RevModPhys.91.015004,jeffries2021small}.  

In recent years, the interplay of processes on microscopic scales with texture formations and the slow dynamics on mesoscopic and macroscopic length and time scales has attracted great interest in many different areas of condensed matter physics, motivating investigations over a very large dynamic range \cite{gardner_high-resolution_2020}. Materials studied comprise soft matter and biological systems, complex bulk compounds, as well as hard condensed matter materials, such as complex metallic alloys, magnets, and superconductors. {\rw For one such material, the ferromagnetic superconductor UGe$_2$, momentum transfers $Q$ down to 0.008 $\AA^{-1}$ and an energy resolution of 1 $\mu$eV was necessary to fully characterize the spin fluctuations \cite{2018Haslbeck}. Such} systems take the form of bulk samples, heterostructures, and nanoscale configurations.

Despite the versatility and heavy demand for SANS measurements, there is essentially no dedicated instrumentation for {\rw SANS} studies {\rw (i.e. pinhole collimation of the incident neutron beam)} of the dynamical properties. On the one hand, triple axis or time-of-flight spectrometers are operated at the limit of their accessible parameter-range without optimized background reduction. {\rw The upper limit of energy (Fourier time) resolution for such techniques is around 600 $\mu$eV (1 picosecond) \cite{bewley2011let}. For backscattering spectrometers the upper limit is 0.3 $\mu$eV ($\sim$2 ns) \cite{gardner_high-resolution_2020}.} On the other hand, conventional neutron spin-echo spectrometers are optimized for large samples and large beam-divergences on the expense of the requirements of SANS. {\rw Optimized in this way they can reach Fourier times of 1000 ns with $Q$ down to 0.001 $\AA^{-1}$ \cite{IN15} and as high as 4 $\AA^{-1}$ \cite{ohl2012spin}. However, being} based on polarized neutrons, conventional spin-echo spectrometers suffer from large incoherent background in studies of hydrogen containing materials, while studies of depolarizing samples (such as magnetic materials) or use of depolarizing sample environment are prohibitively difficult in {\rw all but ideal cases \cite{Farago1986}}.

Magnetic materials is a classification that envelops exceptional diversity, they represent a key material category where spectroscopic studies are a necessary tool for comprehensive exploration and characterization. For example, neutron spectroscopy of the magnetic fluctuations near quantum critical points is a means to extract crucial information about their behavior over broad spatial and dynamical ranges \cite{Mezei1982}. In magnetically frustrated spin systems, often spin-glass transitions can develop resulting in interesting disordered ground states, some of which contain the fingerprints of the Berry phase and other contributions from wavefunction topology. In general, the topological spin textures arising out of careful balances between the magnetic interactions continue to open up new potential cases for applications. Exploring and characterizing the behavior of these types of systems requires the resolution of small momentum and energy transfers, since in general mesoscale textures and microscopic dynamics are involved. {\rw This dual requirement naturally demands a detector with both the highest possible spatial (2D) and temporal resolutions, and ideally the capability to adjust the sample-detector distance for flexibility between the angular coverage and angular resolution. }


A different way to illustrate the need for neutron spin-echo spectroscopy in small angle scattering configuration represent the requests for experimental beam time at the beamline RESEDA at FRM II. With the successful implementation of the longitudinal neutron resonance spin-echo and longitudinal MIEZE technique set-up, an increasing number of experimental proposals has been submitted for studies at momentum transfers below 0.01 {\rw $\AA^{-1}$ - 1 $\AA^{-1}$}, i.e., in the regime of small angle scattering. Notably, since 2015 this concerned over 60\% of the proposals submitted. {\rw Moreover, these} proposals specifically required the MIEZE technique as they were dealing either with hydrogen-containing substances, ferromagnetic or superconducting materials, or strongly depolarizing sample environments. {\rw Examples include (undeuterated) water, Fe, UGe$_2$, and MnSi measurements as summarized in Ref. \cite{2019Franz2}.}

\begin{figure*}
\centering
\includegraphics[width=1\textwidth,clip]{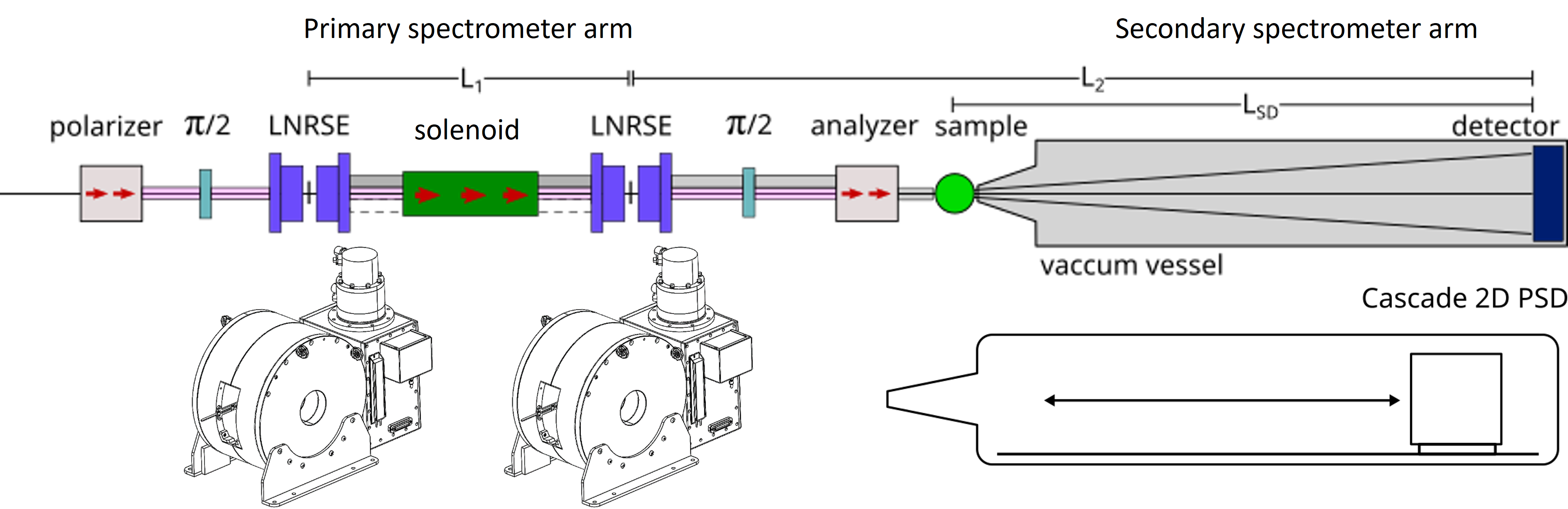}
\caption{Schematic depiction of the RESEDA-MIASANS (Modulation of Intensity Applied to Small Angle Neutron Scattering) set up with sketches of the main new components. The RESEDA beamline is comprised of the primary spectrometer arm with the two resonant spin flippers (left), the sample environment indicated in the center, and secondary spectrometer with the vacuum vessel containing the position sensitive CASCADE detector (right). Shown below the diagram of the beamline components are sketches of the new $B_0$ superconducting solenoids for the RF spin flippers and new detector tank with space for a larger CASCADE detector and on-demand detector translation capability over the full $L_{SD}$ range.  }
\label{fig-rendering}       
\end{figure*}

\subsection{The MIEZE technique}
\label{MIEZEintro}

Neutron spin-echo spectroscopy (NSE) \cite{1972Mezei, 1980Mezei, 2002Keller} allows measurements at both high resolution \emph{and} intensity as compared to triple-axis spectroscopy or time-of-flight spectroscopy, for which high energy resolution can only be reached at the expense of considerable losses of intensity. NSE gets around this tradeoff by producing neutron energy encoding via Larmor precession. In contrast to NSE, in neutron resonant spin-echo (NRSE) {\rw \cite{1999Koppe}} the large magnetic solenoids used to encode the neutron energies in their Larmor precessions are replaced by pairs of resonant spin flippers which are driven by a oscillating radio-frequency magnetic fields {\rw \cite{1992Gaehler,2020Dadisman}}. Note these oscillating RF fields must be surrounded by the appropriate static field $B_0$ tuned to give the Larmor resonance condition. This is essentially the Nuclear Magnetic Resonance (NMR) methodology for neutrons {\rw instead of atomic nuclei}. Resonant spin flippers inherently reduce the differences in magnetic field integral that divergent neutron trajectories experience, and so the complicated field correction coils used in NSE \cite{Farago2015} are not necessary to reach the same instrument performance.  More details about the current status of the NRSE technique and its relationship to other variants of spin-echo spectrometers can be found in Refs. \cite{1987Golub, 2019Franz2}.

{\rw Modulation} of Intensity with Zero Effort (MIEZE) is a neutron resonant spin-echo technique variation where potential information loss via depolarization of the Larmor precessing spins prepared by the resonant flippers is preempted by {\rw placing a spin analyzer just before the sample position, which results in an intensity modulated neutron beam downstream of the analyzer} \cite{1998Besenbock, 2019Franz,2019Jochum, 2019Geerits,oda_tuning_2020,liu_resolution_2020,Dadisman2022}. {\rw No} further beam manipulation components are placed between the sample and the detector in contrast to conventional NSE and NRSE. {\rw The end result is that the same information which was encoded in the Larmor precession (i.e. in the interference between the two spin states of a neutron) is instead encoded in the intensity modulation of the neutron beam. More specifically, before the analyzer the two RF flippers couple together an individual neutron's spin and energy degrees of freedom via resonant spin flips. By selecting just one spin state, the analyzer removes the spin degree of freedom from the neutron. What remains then is a neutron with two separated energy states, prepared in such a way that the separation length goes to zero at the detector position (without any sample interactions). For many neutrons combined, the cumulative consequence of this energy state overlap is an intensity beating with a large amplitude facilitating its observation with a detector.  }

MIEZE requires a position sensitive detector which can accurately measure the time-of-flight of the neutrons \cite{funama_study_2021} after they have scattered from the sample. The detector must be placed at a specific position given by the MIEZE condition, where an oscillating intensity will be observed as determined by the frequencies of the RF flippers {\rw $f_{1}$ and $f_{2}$}:
\begin{equation}
    \frac{L_{1}}{L_{2}}=\frac{f_{2}-f_{1}}{f_{1}}
    \label{eq:MIEZE_condition}
\end{equation}
Here, $L_{1}$ is the distance between the two RF-flippers, whereas $L_{2}$ is the distance between the second RF-flipper and the detector. 

As has been demonstrated a number of times {\rw \cite{Farago1986,2011Haeussler,2018Haslbeck,2019Saubert,Wendl:Master,2019Martin}}, this setup imparts the capability to measure with magnetic samples, magnetic fields in the sample environment, and samples with a high incoherent scattering cross section. These capabilities are acquired without the need for the instrumental complexities faced in ferromagnetic spin-echo \cite{Boucher1985, Farago1986}. Indeed, proof-of-concept measurements under large magnetic fields up to 17 Tesla \cite{2015Kindervater} at RESEDA underscore the inherent advantages of the MIEZE technique as compared to the ferromagnetic spin-echo set-up, where the stray field of the magnet would interfere significantly with the instrument components. {\rw As a result, MIEZE has the highest energy resolution (and $Q$ range) for magnetic scattering \cite{gardner_high-resolution_2020}. }


As spin-echo spectroscopy encodes the {\rw sample} scattering processes in {\rw the spin phase of the Larmor precession induced in the neutrons}, it is inherently sensitive to path length differences of different neutron trajectories \cite{2013Weber}. A major advantage of the longitudinal NRSE and MIEZE set-up concerns the self-correction of path length differences of parallel neutron trajectories, and a strongly reduced sensitivity to differences of divergent beam trajectories \cite{2016Krautloher}. However, the signal in MIEZE studies degrades for large scattering angles due to path length differences, {\rw for which there are currently no feasible compensation methods for the longitudinal configuration of RESEDA. (However, note that in the transverse MIEZE configuration there are multiple viable ways to do such compensation \cite{Dadisman2022,li2022linear})} The strength of this signal reduction has been experimentally quantified for different momentum transfers \cite{2019Franz2} and agrees well with the calculated expectations \cite{webapp}. Fortunately though, all commissioning and scientific studies clearly show {\rw that the} low angle regime is favorable for the MIEZE technique, as flight path differences of neutron trajectories from the sample to the detector \cite{2018Martin} play only a minor role \cite{2005Haeussler, 2019Schober} and the full resolution can be exploited without restrictions on the sample geometry. For this essential reason, the MIEZE technique is especially compatible with SANS type measurements. {\rw This special compatibility has been known for over two decades and demonstrations \cite{bleuel2005misans,bleuel2009experimental,2011Brandl} and experiments \cite{kuhn_time--flight_2021} have been carried out over the years under the MISANS (MI-SANS) designation.}

\subsection{The RESEDA instrument}

RESEDA (REsonance Spin-Echo for Diverse Applications) is a high resolution resonance spin-echo spectrometer located at the cold neutron guide NL5-S in the Neutron Guide Hall West of the FRM II reactor in Garching, Germany {\rw \cite{haussler2007reseda, haussler2008reseda}.} In its present configuration the instrument provides access to a large time and scattering vector range for quasi-elastic measurements \cite{2019Franz, 2019Franz2, 2019Franz3}.

A critical milestone in the development of RESEDA is represented by the development of electronics and coils for the generation of high frequency oscillatory magnetic fields. Using a combination of signal generators, amplifiers, impedance matching transformers, and a carefully designed cascade of tunable capacitors and customized coils, it is possible to achieve a very broad bandwidth from 35\,kHz (lower limit due to the Bloch-Siegert shift \cite{1940Bloch}) to unprecedented 3.6\,MHz \cite{jochum2021oscillatory}, leading to the extremely high resolution of the spectrometer. 

At present in the MIEZE set-up a transmission bender {\rw analyzer} is used in front of the sample for the spin analysis. This transmission bender meets its high specifications, {\rw i.e.} high polarisation ( $>$ 95\%), without affecting the beam properties (such as divergence). However, for small angle scattering geometries, a sufficient degree of collimation is necessary {\rw because} a residual background arises from the roughness of the wafers in the transmission bender \cite{2022TBender}. 
The background may be suppressed by means {\rw of a} 20’ (horizontal and/or vertical) collimation behind the bender. It is important to note that the beam divergence for the small angle scattering geometry is determined by the pinhole collimation on the primary spectrometer arm. In turn, no neutrons are lost with the addition of a 20’ collimator.

\section{MIASANS Objectives}
\label{objectives}

In its previous configuration, RESEDA {\rw was} optimized to cover a wide parameter range, but there is still opportunity {\rw to significantly expand this range in a cost effective manner}. The aim of the RESEDA-MIASANS (longitudinal Modulation of Intensity Applied and optimized for Small Angle Neutron Scattering) project has been to further develop the resonant neutron spin-echo spectrometer RESEDA in the area of small angle scattering {\rw with} higher energy resolution. Two major component upgrades provide these objectives: (i) upgrading the resonant flippers with compact, cryogen free, superconducting magnet coils for generating the static $B_0$ field (ii) {\rw equipping} the secondary spectrometer arm used for MIEZE studies with a detector tank for background reduction and translation stage for increased versatility in positioning the detector and (iii) {\rw reducing the} background from the primary spectrometer arm. 

\begin{figure}
	\centering
	\includegraphics[width=1\columnwidth,clip]{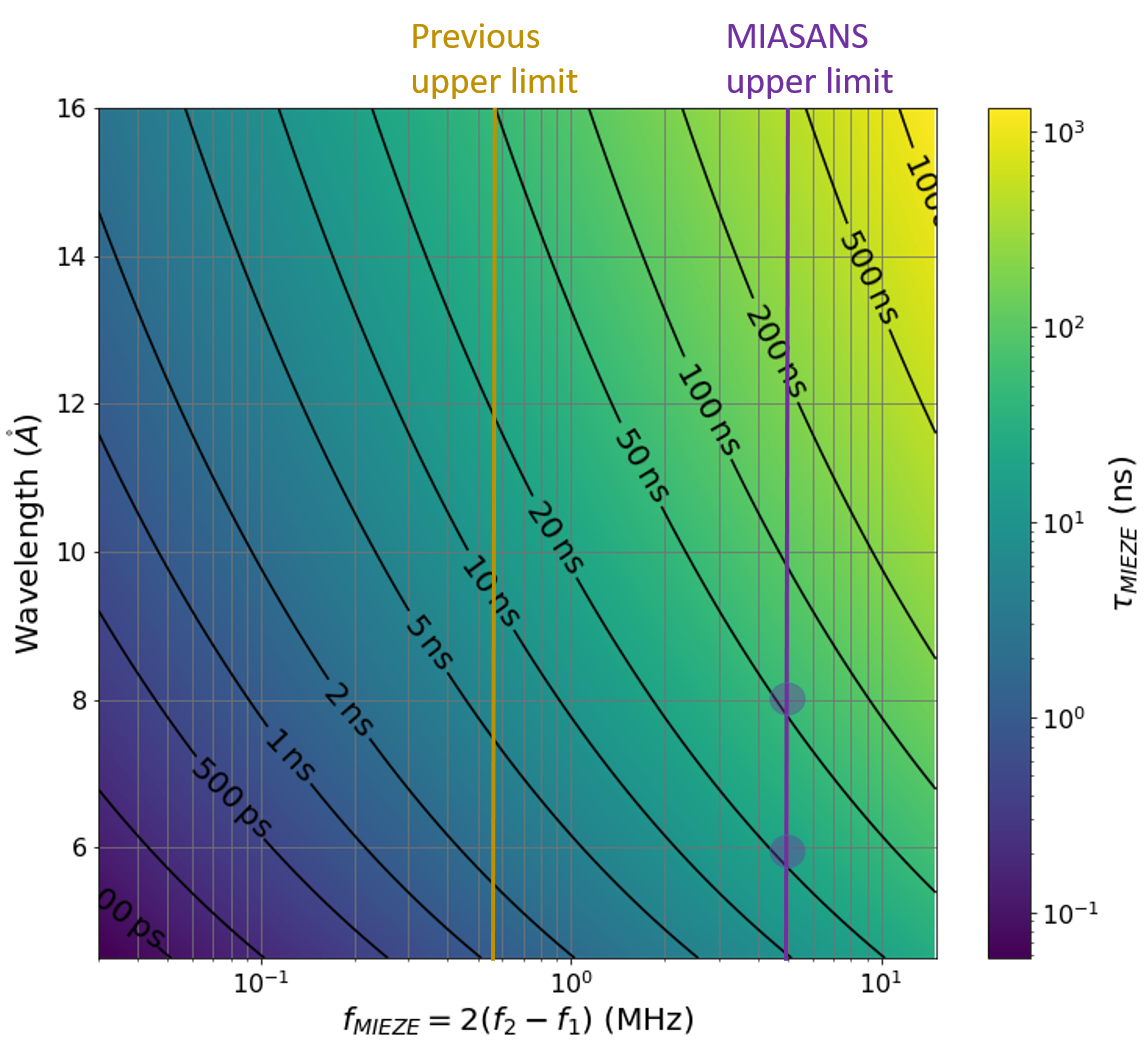}
	\caption{The resolution, i.e. $\tau_{MIEZE}$, of a MIEZE instrument given as a function of neutron wavelength and the frequency difference ($f_{2}-f_{1}$) between the two resonant flippers according to Eq. \ref{eq:echotime}. $\tau_{MIEZE}$ for selected wavelengths (6 $\AA$ is the most common wavelength setting for experiments at RESEDA) are indicated as described in the main text. The newly accessible regimes of $\tau_{MIEZE}$ made possible by the MIASANS project (assuming spin-flips at 8 MHz; $B_0$ = 300 mT) are labeled accordingly. }
	\label{fig-ranges}       
\end{figure}

\begin{figure}
	\centering
	\includegraphics[width=1\columnwidth,clip]{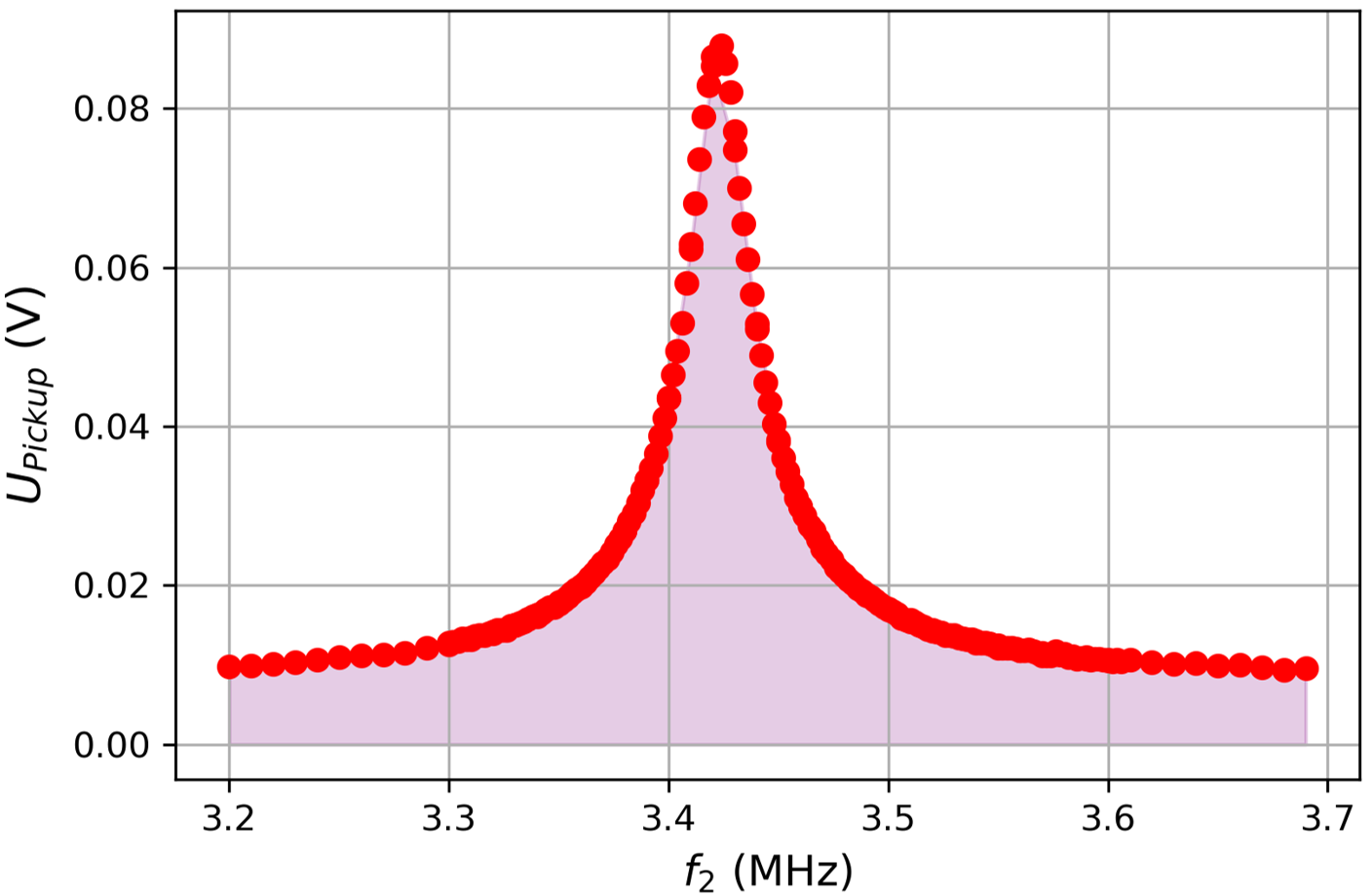}
	\caption{{\rw A representative} resonance scan (as described in Ref. \cite{jochum2021oscillatory}) of the second MIEZE RF flipper coil {\rw showing a peak at 3.42 MHz} using the newly installed superconducting coils which provide the static field $B_0$ around the resonance coils for longitudinal resonant spin-echo. As shown in Fig. 9 of Ref. \cite{jochum2021oscillatory}, these new magnets {\rw will} allow for resonant flips at a record high frequency $f$ = 3.6 MHz at RESEDA.}
	\label{fig-resonance}       
\end{figure}

A schematic depiction of the RESEDA instrument with the new MIASANS components is shown in Fig. \ref{fig-rendering}. Note that RESEDA is {\rw a modular instrument} and development of this tank will not affect the possibility to perform {\rw longitundinal} NRSE measurements.  In addition, it is crucial to further optimize {\rw the neutron} beam generated on the primary spectrometer arm. As introduced in detail in Sec. \ref{MIEZEintro} above, in a MIEZE spectrometer all spin manipulations are carried out on the primary spectrometer arm. For high scattering angles and relaxed resolutions, the neutron flux on the sample is maximised by means of m = 1.2 supermirror guides. However, for small angle scattering and high resolution this would severely limit the accessible momentum-range and momentum resolution. Even with a pinhole collimation, small angle scattering from the aluminum in the RF-flippers and the air surrounding them in the space outside the neutron guides creates a significant amount of unwanted background. An objective in the MIASANS project is to quantify this background and implement the most effective ways to mitigate it.

\section{Primary spectrometer arm}

\subsection{Superconducting static field coils}
\label{SCcoils}

Here we briefly describe the significantly increased resolution, i.e. $\tau_{MIEZE}$, capabilities from higher frequency spin flips generated in the two resonant RF flippers on the primary arm of RESEDA. Fig. \ref{fig-ranges} illustrates the increase in the upper limits of $\tau_{MIEZE}$ as the MIEZE frequency is raised. Analogous to the so-called spin-echo time, $\tau_{MIEZE}$ depends on the neutron wavelength $\lambda$, the frequency difference between the two RF flippers {\rw ($f_{2}-f_{1}$)}, and the sample-detector distance $L_{SD}$:
\begin{equation}
    \tau_{MIEZE}=\frac{m^{2}}{h^{2}}\cdot\lambda^{3}\cdot2\pi(f_{2}-f_{1}) \cdot L_{SD}
    \label{eq:echotime}
\end{equation}
($m$ = neutron mass, $h$ = Planck's constant). The substantial improvement of accessible $\tau_{MIEZE}$ is made possible with the newly procured superconducting solenoids which provide static $B_0$ fields up to 300 mT with the required field homogeneity over the active resonant flipping areas.

To reach the maximum resolution and full dynamic range \cite{2019Jochum} of the existing MIEZE set-up at RESEDA, RF spin flips at frequencies up to 4 MHz are required \cite{jochum2021oscillatory}. However, the maximum flipping frequency was previously limited to roughly 1 MHz due to capacitive stray coupling between the the RF coils and conventional Helmholtz coils used to generate the static $B_0$ field. This was attributed to capacitive coupling causing a parasitic damping of the RF oscillations, and as such this situation could be resolved by reducing/eliminating such couplings. 

Indeed, this solution was demonstrated using a small high-T$_C$ superconducting magnet of the sample environment group at FRM II to generate the static $B_0$ field, and with that the resonant flip at a frequency of 3.6 MHz using a neutron wavelength of 6 $\AA$ was confirmed as shown in Fig. 9 of Ref. \cite{jochum2021oscillatory}. For this test a static field $B_0$ = 122 mT was required. 

Therefore, two new 300mT high temperature superconductor (HTS) magnet systems were designed 
together with the company HTS-110 (based in New Zealand) to the specifications achieved in the aforementioned test-set up. Following successful performance characterization, {\rw these} new customized superconducting magnets from HTS-110 were integrated on the RESEDA beamline. 
The first functional tests of these magnets show resonant frequencies up to {\rw 3.42} MHz, as {\rw seen} in Fig. \ref{fig-resonance}. {\rw Future test with neutrons on this new RESEDA configuration are expected to reproduce the data shown in Fig 9 of Ref. \cite{jochum2021oscillatory} }

The newly installed superconducting $B_0$ coils have been shown to provide field strengths up to 320 mT. This static field strength will permit resonant spin-flips up to 8 MHz on the primary arm of RESEDA, which corresponds to maximum {\rw $f_{MIEZE}$ = 2($f_{2}-f_{1}$)} of 4.9 MHz. This enables $\tau_{MIEZE}$ {\rw for practical experiments} to reach {\rw ~20 and ~50 nanoseconds} (with $L_{SD}$ = 3340 mm), for neutron wavelengths of {\rw 6 and 8 $\AA$} respectively as can be seen in Fig. \ref{fig-ranges}. {\rw Using wavelengths above 8 $\AA$ to reach higher $\tau_{MIEZE}$ is possible in principle, but impractical due to the substantial loss of neutron flux at those wavelengths.} With these improvements in the field integral ($\sim$0.5 Tm) and $\tau_{MIEZE}$ at RESEDA, the instrument makes a considerable step closer to the performance of world leading NSE spectrometers such as J-NSE Phoenix \cite{2019Pasini} and IN15 \cite{IN15}. 


Thus in the future it is planned to construct and install compact resonant spin-flipping solenoids which can go up to 8 MHz instead of the 3.6 MHz \cite{jochum2021oscillatory}. At such high frequencies the density of phase rings on the detector will increase. In order to extract meaningful results from measurements, the phase difference contained within one detector pixel should be less than 20$^{\circ}$ \cite{2011Brandl, 2018Martin}. For the case of a MIEZE frequency {\rw $f_{MIEZE}$ =} 4.9 MHz (and time resolution of 20 MHz using the 4 point method \cite{4point_echo}), then on the outermost edge of the detector where the phase rings are densest there will be a phase difference of 115$^{\circ}$ within one pixel. Such an extreme phase difference in one pixel would result in severe reduction in the measured signal (contrast). {\rw In order to} try and avoid this in such extreme cases, the area of the individual detector pixels needs to be {\rw further} reduced, which may be feasible in the next iterations of the $^{10}$B GEM foil production {\rw for CASCADE detectors \cite{2016Koehli,kohli2019cascade}}. 

\subsection{Background reduction}
\label{background_reduction}

A substantial source of background scattering at the RESEDA instrument has been identified to originate from the air in front of the sample location. This scattering from air may be reduced by customized evacuated adapters matched to the different sample environments available, namely the closed cycle cryostats, furnaces, and 5~T and 12~T superconducting magnets. Using a system of standard KF flanges in combination with sapphire windows, {\rw a considerable reduction of the air within the neutron flight paths has been achieved without compromising} the full range of required distances and geometries {\rw for SANS}. However, there are still several portions of the primary arm where small volumes of air cut through the neutron paths, such as gaps in between the {\rw different} evacuated neutron guides {\rw to accomodate} the resonant flippers and other vital components. Further gains in background reduction could be achieved by minimizing these portions of air. 

In addition to the air, another source of SANS background originates from the material of the longitudinal RF coils, which consist of pure aluminium with a thickness of 0.2 mm. It is noteworthy that this is significantly less material in the path of the neutron beam {\rw in comparison to} the transverse NRSE configuration, where the $B_0$ coil has a thickness of several mm (in the previous TNRSE design at RESEDA, this thickness was 6 mm plus a layer of anodized aluminium). Replacing the anodized aluminium with pure aluminium alone reduced the background scattering produced by the coils by a factor of 100 \cite{kindervaterThesis}. In the current configuration of RESEDA, no transmission losses {\rw through the RF coils} have been observed in {\rw either} McStas simulations or experiments.

In an attempt to distinguish the contributions from these  different sources of background {\rw arising within RESEDA's} primary arm {\rw with pinhole slit apertures configured}, we {\rw utilized} the RESEDA instrument file setup in the Monte Carlo ray tracing program McStas \cite{willendrup_mcstas_2020, willendrup_mcstas_2021}. With the information from these simulations, the theoretical configurations for optimized SANS background reduction may be {\rw discerned}. 

\begin{figure}
	\centering
	\includegraphics[width=1\columnwidth,clip]{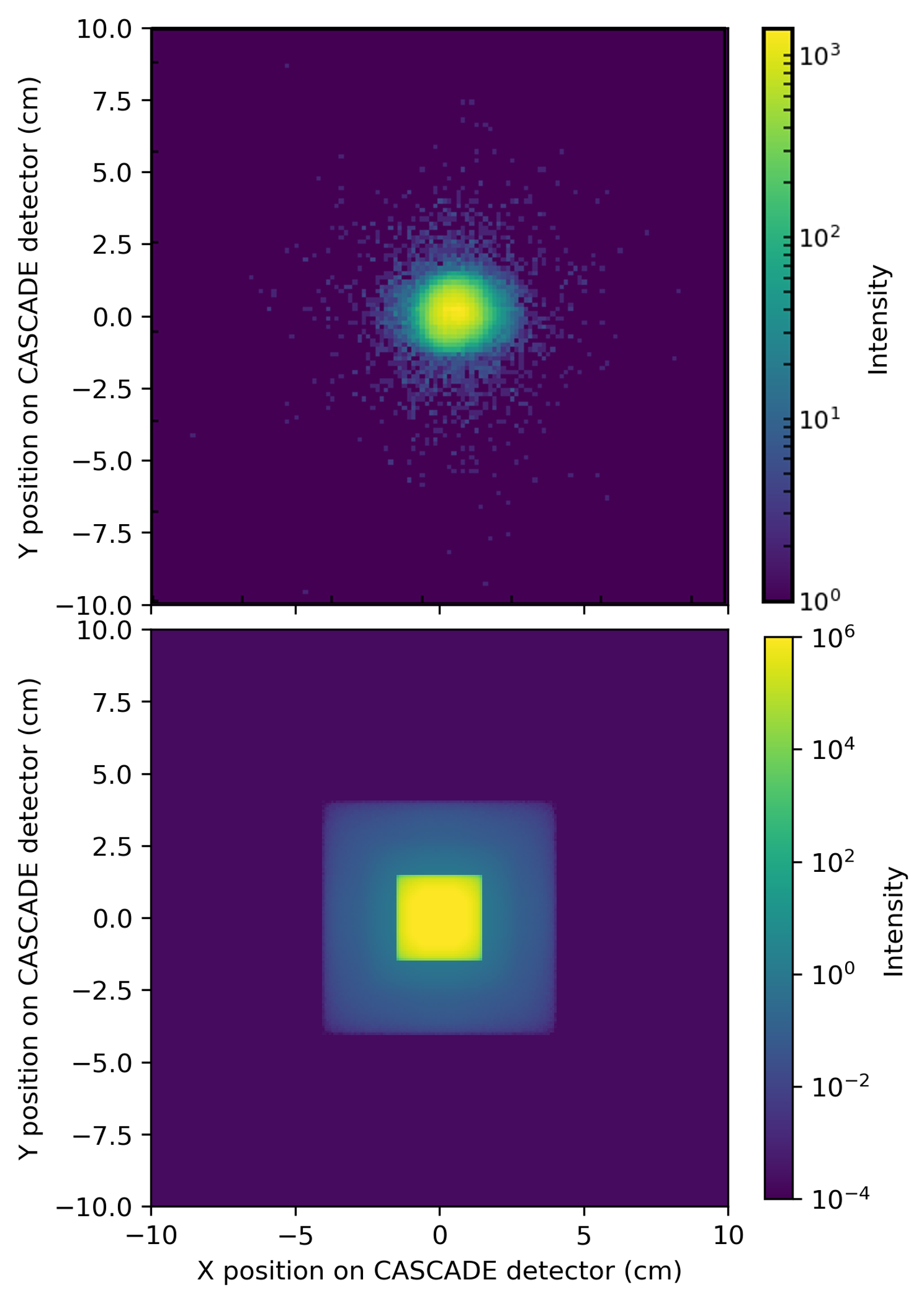}
	\caption{Typical detector image from the 2D position sensitive CASCADE detector \cite{2016Koehli,kohli2019cascade} at RESEDA, illustrating the small angle scattering scattering background halo surrounding the direct beam. (top) Experimentally measured detector image, with standard 20' horizontal and vertical collimation. (bottom) McStas simulation of the same configuration, without any air scattering being simulated, showing that the background from SANS scattering at the RF flipper components accounts for some of this halo. This can be qualitatively compared with the SANS measurements from pure aluminum RF coils shown in Fig 2.5 of Ref. \cite{kindervaterThesis}. {\rw Note that because diffraction and air scattering are not accounted for in the McStas simulation, the square shape from the guides and apertures is traced through to the detector image.}}
	\label{fig-sim2d}       
\end{figure}

As shown in Fig. \ref{fig-sim2d}, {\rw a halo of} background around the direct beam can be observed in McStas simulations, where it is possible to easily insert and remove the background sources. Based on such simulations, it can be deduced that the majority of this background is incoherent scattering originating from air and aluminum components in the neutron flight path on the primary arm of RESEDA. Thus, steps will be taken to reduce these remaining portions of air on the primary arm. Importantly, the McStas simulations allowed us to test the hypothesis that the reflections {\rw carried through the m = 1.2 neutron guides} may enhance the background halo around the direct beam. Simulation {\rw results} with an m = 0 guide {\rw were negligibly different in comparsion with} the m = 1.2 guide. 

\section{Secondary spectrometer arm}

\subsection{Vacuum Vessel}
\label{vacuumtank}

The secondary spectrometer arm currently available at RESEDA for the MIEZE set-up, which consists of an evacuated flight tube (only 30 cm in diameter) with the CASCADE detector attached to the outside of the far end, 
requires massive improvements for optimization of SANS measurements. The main reasons such an upgrade is necessary are {\rw (i)} the reduction in the background scattering in SANS geometry and {\rw (ii)} the facilitation of maximum flexibility between the largest angular coverage with relaxed {\rw time} resolutions and reduced angular coverage for the {\rw highest $\tau_{MIEZE}$}. For the purpose of this flexibility, it is mandatory to mount the detector inside a {\rw vacuum tank large enough to accommodate a translation} rail system akin a conventional SANS detector tank. In addition, mounted in front of the detector will be three customized beamstops, which can be {\rw horizontally adjusted} on-demand anywhere along the active detector area {\rw at the height of the direct beam}. {\rw These automated beamstops will save valuable experiment time over the manual manner in which this direct beam attenuation was done before.} 


The new 101 cm diameter vacuum vessel, manufactured by the Added Value Solutions (AVS) firm, is configured to accommodate a position sensitive {\rw CASCADE} detector \cite{2016Koehli, kohli2019cascade} covering an area of 30 x 30 cm$^2$ expected to be commercially available in the coming years. This will enable potential increases of the count rate by a factor of two as compared to the current 20 x 20 cm$^2$ {\rw detectors discussed in the following section}.


\subsection{CASCADE detector}
\label{CASCADE}
The existing 20 x 20 cm$^2$ CASCADE detector \cite{2016Koehli,kohli2019cascade} at RESEDA has proven very effective {\rw \cite{2011Haeussler} especially since the foil waviness has been fully characterized \cite{2019Franz}}. Such waviness can be corrected in the standard data reduction procedure (MIEZEPY \cite{2019Schober}). Spin-echo signals up to 100 ns have been recorded in the direct beam, {\rw albeit with very low intensity}. 

As noted previously in Sec. \ref{SCcoils}, phase effects across the pixels will eventually become one of the resolution bottlenecks of the (flat) CASCADE detector. 
The 1.56 mm pixel size of the CASCADE detector is (much) smaller than the average pixel size of a typical SANS detector (8 mm), and therefore any effects due to the pixelated nature of the detector {\rw have not yet been observed in experiments}. For example, given a MIEZE frequency of 2.5 MHz (time resolution of 10 MHz using the 4 point method \cite{4point_echo}), and using 6 $\AA$ neutrons with $L_{SD}$ = 3340 mm, one estimates a phase variation in a single pixel of $\sim$15$^{\circ}$. This value is still below the threshold of 20$^{\circ}$ given in \cite{2011Brandl, 2018Martin}. {\rw Even though the neutron counts in each individual small pixel will be quite low, with the proper phase correction procedure implemented with the MIEZEPY software \cite{2019Schober}, the cumulative intensity of a custom group (bin) of pixels can be determined.}

The setting accuracy and stability of the power supplies yields a sufficient phase stability for the frequencies used so far at RESEDA (i.e. the phase does not drift significantly over time when the settings are maintained). Furthermore, all frequency generators and RF circuits receive their signal from the same high precision Rubidium atomic clock time-standard. However, in this context, other environmental contributions to the phase stability need to be considered, {\rw particularly effects from temperature variations}. Based on the thermal expansion of alumninum, a 1~K temperature variation with the aforementioned parameters induces a $\Delta L_{SD}$ and a corresponding phase shift below the critical 20$^{\circ}$ threshold. {\rw This however does further motivate a reduction in detector pixel size in order to better account for temperature variations at the highest MIEZE frequencies.}
For these reasons, once the time resolution of the CASCADE detector is pushed beyond 10 MHz, it will become necessary to improve upon the spatial resolution {\rw for use at the upper resolution limits of the MIEZE technique}. Fortunately, recent innovations in the GEM foil production processes \cite{shah2019impact} will allow CASCADE-type detectors with a larger active area {\rw and smaller pixel size} to be manufactured, making such improvements in spatial resolution feasible \cite{2011Brandl, 2018Martin}.

CDT CASCADE Detector Technologies has completed the design and construction of a detector housing {\rw large enough to fit} future $^{10}$B GEM detectors. 
The housing is equipped with a {\rw circular} detector window of 44 cm diameter {\rw which surrounds a} 30 x 30 cm$^2$ square active detector area. The detector window thickness is minimized to avoid {\rw secondary scattering effects as much as possible}. Furthermore, the new detector {\rw is} adapted to minimize local power dissipation such that active cooling {\rw of the detector electronics} is unnecessary. Initially this new 20 x 20 cm$^2$ CASCADE detector will be placed in this new housing, for use at RESEDA until {\rw the aforementioned} 30 x 30 cm$^2$ detector with increased spatial resolution becomes available. \\

\section{Summary}
We have designed a MIEZE set-up as part of the
spectrometer RESEDA at FRM II that is optimized for the requirements of small angle neutron scattering (MIASANS). This project builds upon major progress with the longitudinal neutron
resonance spin-echo method, achieved in recent years at RESEDA \cite{2016Krautloher, 2019Franz, 2019Franz2, 2019Franz3, 2019Jochum, jochum2021oscillatory, 4point_echo}. We have so far {\rw demonstrated} resonant spin flips up to 3.6 MHz, and {\rw established that our CASCADE detector is sufficiently sensitive to the highest frequency intensity modulations currently available}. As a crucial milestone, the superconducting magnet systems from HTS-110 were delivered and successfully put into operation at the RESEDA beamline. 

The results of the Monte Carlo simulations indicate substantial contributions from scattering by air and aluminum components on the primary arm of RESEDA in front of the sample environment. As the amount of aluminum in the neutron beam is already minimized, further steps will be taken to minimize the {\rw volume} of air in the neutron flight paths. 

Another important milestone was the design of the detector tank, which is currently in production by AVS. In comparison to standard SANS beamlines, the new RESEDA detector tank will be mounted onto a reconstructed support {\rw structure for} the MIEZE spectrometer arm, permitting continued studies at large scattering angles with relaxed resolutions in addition to SANS measurements with high resolutions. As an additional feature needed to attenuate the direct beam, three customized movable beam stops will be installed. 

Finally, {\rw a newly enhanced CASCADE detector covering an area of 20 x 20 cm$^2$ and its vacuum housing has been manufactured by CDT. The vacuum housing is large enough for} a detector with an active area of 30 x 30 cm$^2$ expected to be commercially available in the {\rw not-to-distant} future. This will result in increases of the count rate by a factor of two under optimal conditions as compared to the current set-up {\rw of the RESEDA instrument.}\\
\\


\section{Acknowledgments}
We emphatically thank P. B\"oni, C. Fuchs, K. Fellner, P. Bender, P. Wild, L. Vogl, M. Kleinhans, N. Huber, A. Wendl, the E51 group in the Physics Department of the Technical University of Munich, and the Forschungs-Neutronenquelle Heinz Maier-Leibnitz (FRM II) for support, assistance with setup and testing of the equipment described here, and fruitful discussions. We also wish to thank T. Huang from HTS-110, P. Noguera Crespo and F. Cacho-Nerin from Added Value Solutions (AVS), and M. Klein and C. Schmidt from CDT technologies GmBH for their support in manufacturing the MIASANS equipment described here. This project has been funded by the by the German Bundesministerium für Bildung und Forschung (BMBF, Federal Ministry for Education and Research) through the  Project No.\ 05K19W05 (Resonante Longitudinale \mbox{MIASANS} Spin-Echo Spektroskopie an RESEDA). We also acknowledge support of the German excellence initiative under EXC-2111 (MCQST, project-id 390814868).\\


%
 \bibliography{citations}
%
%
%
%

\end{document}